\documentclass[twocolumn,showpacs,preprintnumbers,amsmath,amssymb]{revtex4}
\usepackage{graphicx}
\usepackage{dcolumn}
\usepackage{bm}

\newcommand{\bef}{\begin{figure}}
\newcommand{\eef}{\end{figure}}

\newcommand{\be}{\begin{equation}}
\newcommand{\ee}{\end{equation}}
\newcommand{\bea}{\begin{eqnarray}}
\newcommand{\eea}{\end{eqnarray}}
\begin{document}

\title{Using the $\phi$-meson elliptic flow to map the strength of partonic interaction}

\author{Md. Nasim}
\affiliation{National Institute of Science Education and Research, Bhubaneswar-751005, India }

\begin{abstract}
 A compilation of recently measured STAR data for elliptic
flow ($v_{2}$) of  $\phi$ mesons in RHIC Beam Energy Scan program and
comparison with a multiphase transport model (AMPT) has been
presented. The experimental data at $\sqrt{s_{NN}}$ $\geq$ 19.6 GeV
agrees well with string melting version of the AMPT model. The model
includes partonic interactions and quark coalescence as a mechanism of
hadronization.  
This indicates that there is a  substantial
contribution to collectivity from partonic interactions at 
$\sqrt{s_{NN}}$ $\geq$ 19.6 GeV.
The  measured $\phi$-meson $v_{2}$  at  $\sqrt{s_{NN}}$ = 7.7 and 11.5
GeV are found to be smaller than those obtained from AMPT model
without partonic interactions. This indicates
negligible contribution of partonic collectivity to the observed
$\phi$-meson $v_{2}$ at $\sqrt{s_{NN}}$ $\leq$ 11.5 GeV.
\end{abstract}
\pacs{25.75.Ld}
\maketitle

\section{INTRODUCTION}
One of the main goals of the high energy heavy-ion collision experiments is to study the various aspects of the
QCD phase diagram~\cite{whitepapers}. With this purpose the Relativistic Heavy Ion
Collider (RHIC) has finished the first phase of the Beam Energy Scan (BES)
program~\cite{bes_res1,v2_BES_prc,v2_BES_prl}. The aim of the BES program was to look for changes in
observation of various measurements as a function of beam energy to
establish transition region between the partonic and/or hadronic
dominant interactions in the QCD phase diagram~\cite{bes_moti}.\\
The elliptic flow parameter $v_{2}$ is a good tool for studying
the system formed in the early stages of high energy collisions
at RHIC~\cite{hydro,hydro1,hydro2,hydro3,hydro4}. It describes the azimuthal momentum anisotropy of particle
emission in heavy-ion collisions. It is defined as the
second harmonic coefficient of the azimuthal Fourier decomposition of
the momentum distribution with respect to the reaction plane angle
($\Psi$) and can be written as
\begin{equation}
v_{2}=\langle\cos(2(\phi-\Psi))\rangle,
\end{equation}
where $\phi$ is emission azimuthal angle~\cite{method}. 
According to hydrodynamical description $v_{2}$ is an early time phenomenon and  sensitive to the
equation of state of the system formed in the collision~\cite{hydro,hydro1,hydro2,hydro3,hydro4,early_v2}.
The results from Relativistic Heavy Ion Collider (RHIC) on $v_{2} $ as a
function of transverse momentum ($p_{T}$)  shows
that at low $p_{T}$ elliptic flow of identified
hadrons follows mass ordering (lower $v_{2}$ for heavier hadrons than that of
lighter hadrons) whereas at intermediate $p_{T}$ all
mesons and all baryons form  two different groups.
When $v_{2}$ and $p_{T}$ are scaled by number of constituent
quarks of the hadrons, the measured $v_{2}$ values are consistent with
each other as the parton coalescence or recombination models
predicted~\cite{ncq1,ncq1a,ncq_phi}. This observation, is known as
 number of constituents quark scaling (NCQ scaling). This effect has been interpreted as
 collectivity being developed at the partonic stage of the evolution
 of the system in heavy-ion collision~\cite{ncq2}. \\ Although the parton coalescence
or recombination model can successfully explain the observed quark
scaling in experimental data but one can not say that only NCQ scaling of
identified hadrons~\cite{ncq2}  is sufficient signature for the
formation of de-confined matter. The study of NCQ scaling of
identified hadrons from UrQMD model shows that the pure hadronic medium
can also  reproduced such scaling in
$v_{2}$~\cite{ncq_urqmd1,ncq_urqmd2,ncq_urqmd3}.
This is due to modification of
initially developed $v_{2}$ by later stage hadronic interactions ~\cite{ncq_urqmd2}. So the $v_2$ of those particles which do not interact with
hadronic interaction will be the clean and good probe for early
dynamics in heavy-ion collisions.  The $\phi$ meson, which is the
bound state of $s$ and $\bar{s}$ quark, has small interaction cross-section
with other  hadrons~\cite{smallx} and freezes out early~\cite{whitepapers}. Due to small hadronic interaction cross-section,  $\phi$-mesons $v_{2}$
are almost unaffected by later stage interaction and it will have negligible value if $\phi$ mesons are not
produced via $s$ and $\bar{s}$ quark
coalescence~\cite{NBN,BN}. Therefore, it is very important to study the $\phi$-meson $v_{2}$ in
BES program at RHIC.

The paper is organized in the following way. In the section \textrm{II},  AMPT
model has been briefly discussed. Section \textrm{III} describe the comparison of experimentally measured $\phi$-meson
$v_{2}$ with the corresponding results from the AMPT model (version
1.11). Finally the summary and conclusion has been discussed in
section \textrm{IV}.
\begin{figure*}
\includegraphics[scale=0.6]{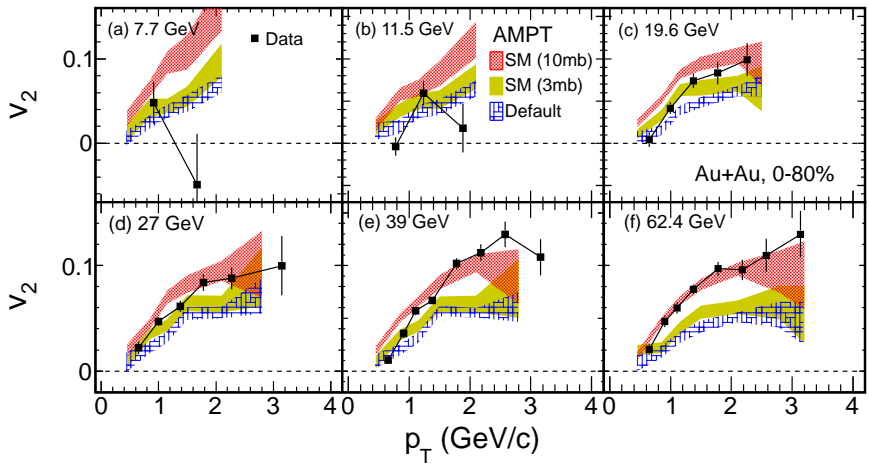}
\caption{(Color online) The $\phi$-meson $v_{2}(p_{T})$ for Au+Au
  minimum-bias collisions at mid-rapidity($|\it y|$ $<$ 1.0 ) from
the STAR experiment at RHIC compared to the corresponding
AMPT model calculation at various beam energies~\cite{v2_BES_prc}. The errors shown are statistical.}
\label{v2_pt}
\end{figure*} 
\section{The AMPT Model}
The AMPT model, which is a hybrid transport model, has four main stages: the initial conditions,
partonic interactions, the conversion from the partonic to
the hadronic matter, and hadronic interactions~\cite{ampt}. It uses the same initial conditions from HIJING~\cite{hijing}.
Scattering among partons are modelled by Zhang’s parton cascade~\cite{ZPC}, which calculates two-body parton scatterings
using cross sections from pQCD with screening masses. In the default
AMPT model, partons are recombined
with their parent strings and when they stop interacting,
the resulting strings fragment into hadrons according to the Lund
string fragmentation model~\cite{lund}. However
in the string melting scenario (labeled as AMPT-SM), these strings are converted
to soft partons and a quark coalescence
model is used to combine parton into hadrons. The
evolution dynamics of the hadronic matter is described
by A Relativistic Transport (ART) model.  The interactions between the minijet partons in the AMPT Default model and those between partons in the
AMPT-SM could give rise to substantial $v_2$. Therefore, agreement between the data
and the results from AMPT-SM would indicate the contribution of
partonic interactions to the measured $v_2$. The parton-parton interaction cross section
in the string-melting version of the AMPT is taken to be  3mb and 10
mb. In this study,
approximately 1.5 million events for each configuration were
generated for minimum-bias Au+Au collisions.\\
\section{Results and Discussion}
In this section, the $\phi$-meson $v_{2}$ measured by STAR experiment
at mid-rapidity ($|y| < $ 1.0) for $\sqrt{s_{NN}}$ = 7.7 - 200 GeV~\cite{v2_BES_prc,v2_BES_prl} has been compared with AMPT model. 
$\phi$ mesons are identified from the $K^{+}$ and  $K^{-}$ decay
channel, the same method as used in experimental analysis.\\
\subsection{Differential $v_{2}$} 
Figure~\ref{v2_pt} shows the comparison of elliptic flow of  $\phi$ mesons in 0-80$\%$
minimum-bias Au+Au collisions at mid-rapidity ($|y| < $ 1.0) for $\sqrt{s_{NN}}$ = 7.7,
11.5, 19.6, 27, 39 and 62.4 GeV with the corresponding results from
the AMPT model~\cite{v2_BES_prc,v2_BES_prl}. The measured data points are
compared with both AMPT String Melting (3 mb and 10 mb parton-parton
cross-section) and AMPT Default version. 
At $\sqrt{s_{NN}}$ = 62.4 GeV experimental data are in a good
agreement with AMPT String Melting model with 10 mb parton-parton
cross-section. This is also true for $\sqrt{s_{NN}}$ = 200 GeV as
reported in Ref.~\cite{NBN}. The measured $\phi$ $v_{2}$, for $p_{T} <
1.5$ GeV/c, lie between 3 mb and 10 mb
 for the energy range $ 19.6$ $\leq$ $\sqrt{s_{NN}}$ $\leq$ 39
 GeV, but in order to explain the measurements for $p_{T} > 1.5$ GeV/c,
 a parton-parton cross-section of the oder of 10 mb is required. 
None of the above model can  explain the trend of
$\phi$-mesons $v_{2}$ at $\sqrt{s_{NN}}$ = 7.7 and 11.5 GeV where the
event statistics for data is also small.  As we expect that the $\phi$-meson $v_{2}$ mostly reflect the
collectivity from the partonic phase, therefore from the comparison
of  experimental data with AMPT model one can conclude that  the partonic collectivity has been
developed for $\sqrt{s_{NN}}$ $\geq$ 19.6 GeV at RHIC. Whereas the contribution from the partonic
collectivity to the final collectivity seems  negligible at
$\sqrt{s_{NN}}$ $\leq$ 11.5 GeV.

\subsection{$p_{T}$ Integrated Elliptic Flow ($\langle v_{2} \rangle$)}
\begin{figure}
\begin{center}
\centerline{\includegraphics[scale=0.45]{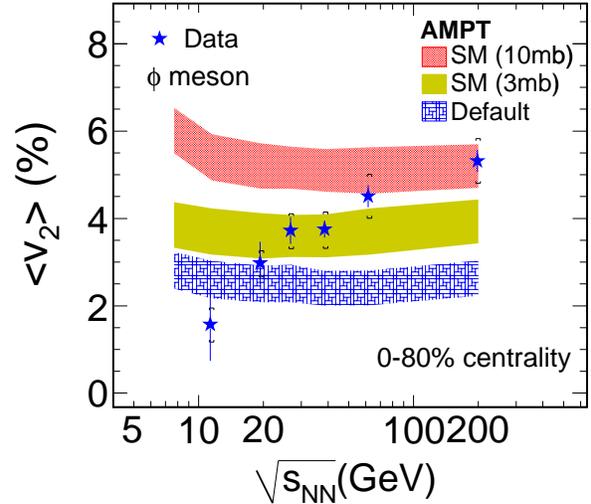}}
\caption{(Color online) The $p_T$ integrated $\phi$-meson $v_{2}$ for Au+Au
  minimum-bias collisions at mid-rapidity($|\it y|$ $<$ 1.0 ) from
the STAR experiment at RHIC are compared to the corresponding
AMPT model calculation at various beam energies. Systematic errors are
shown by cap symbol on experimental data.}
\label{avg_v2}
\end{center}
\end{figure}

The $p_{T}$ integrated elliptic flow $\langle v_{2}\rangle$, which is also an
interesting observable, can be defined as:
 \begin{equation}
  \langle v_{2} \rangle = \frac{\int v_{2}(p_{T})dN/dp_{T}dp_{T}}{\int dN/dp_{T}dp_{T}},
\label{avg_v2_eq}
\end{equation}
i.e. the $\langle v_{2}\rangle$ folds the measured $v_{2}$ versus
$p_{T}$ with the $p_{T}$ distribution ($dN/dp_{T}$) of that
particle.
\begin{figure}
\begin{center}
\includegraphics[scale=0.35]{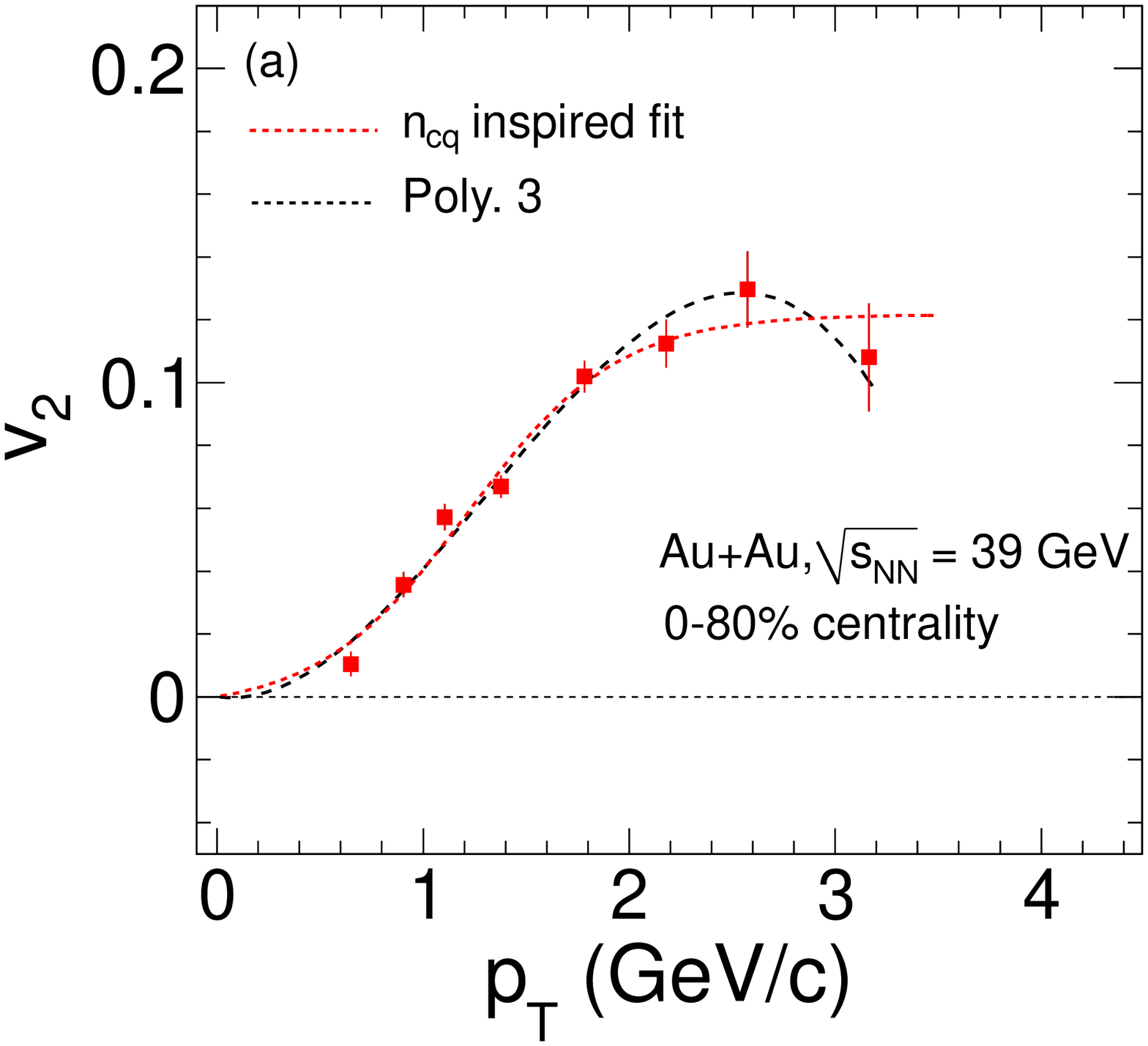}
\includegraphics[scale=0.35]{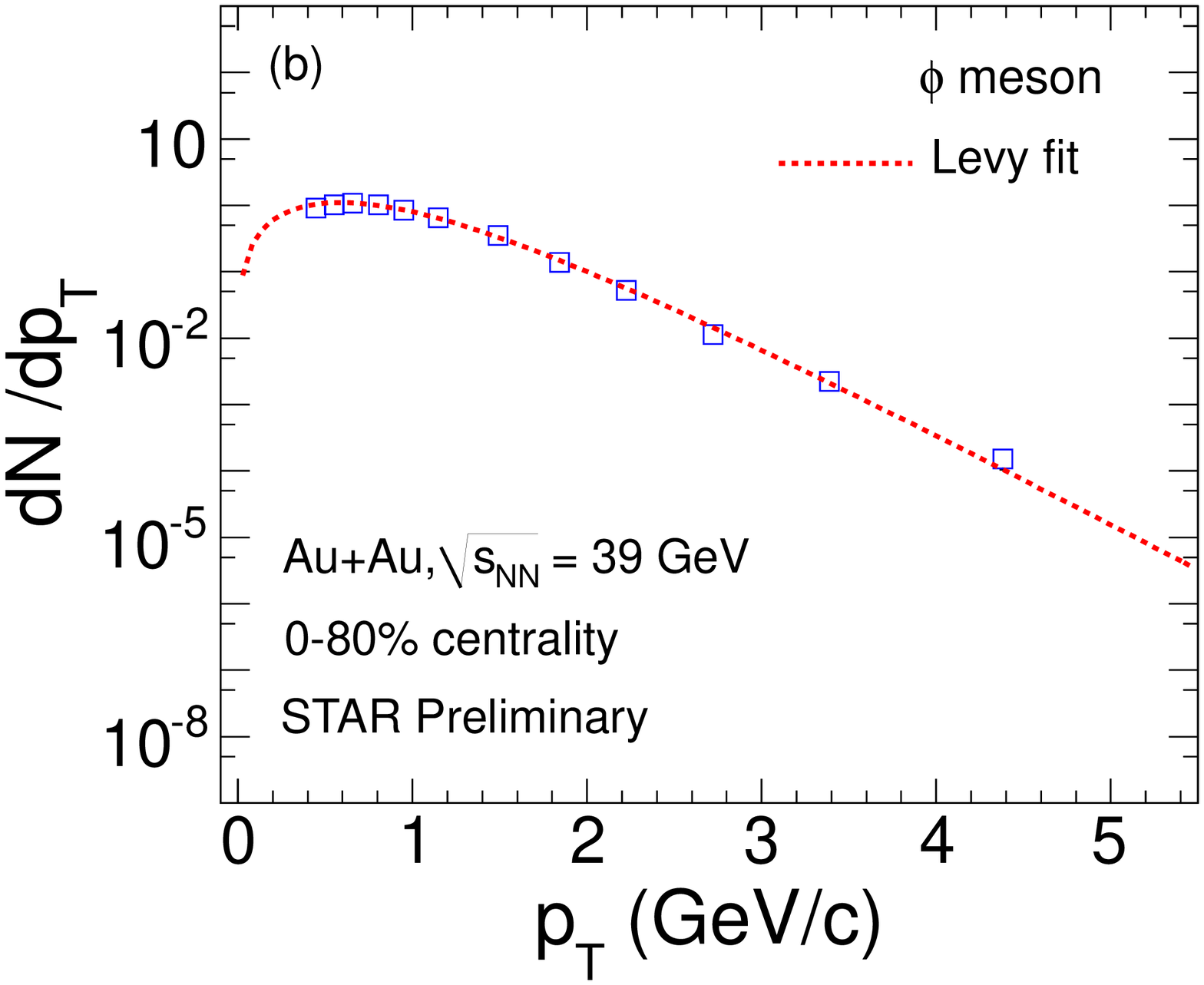}
\includegraphics[scale=0.35]{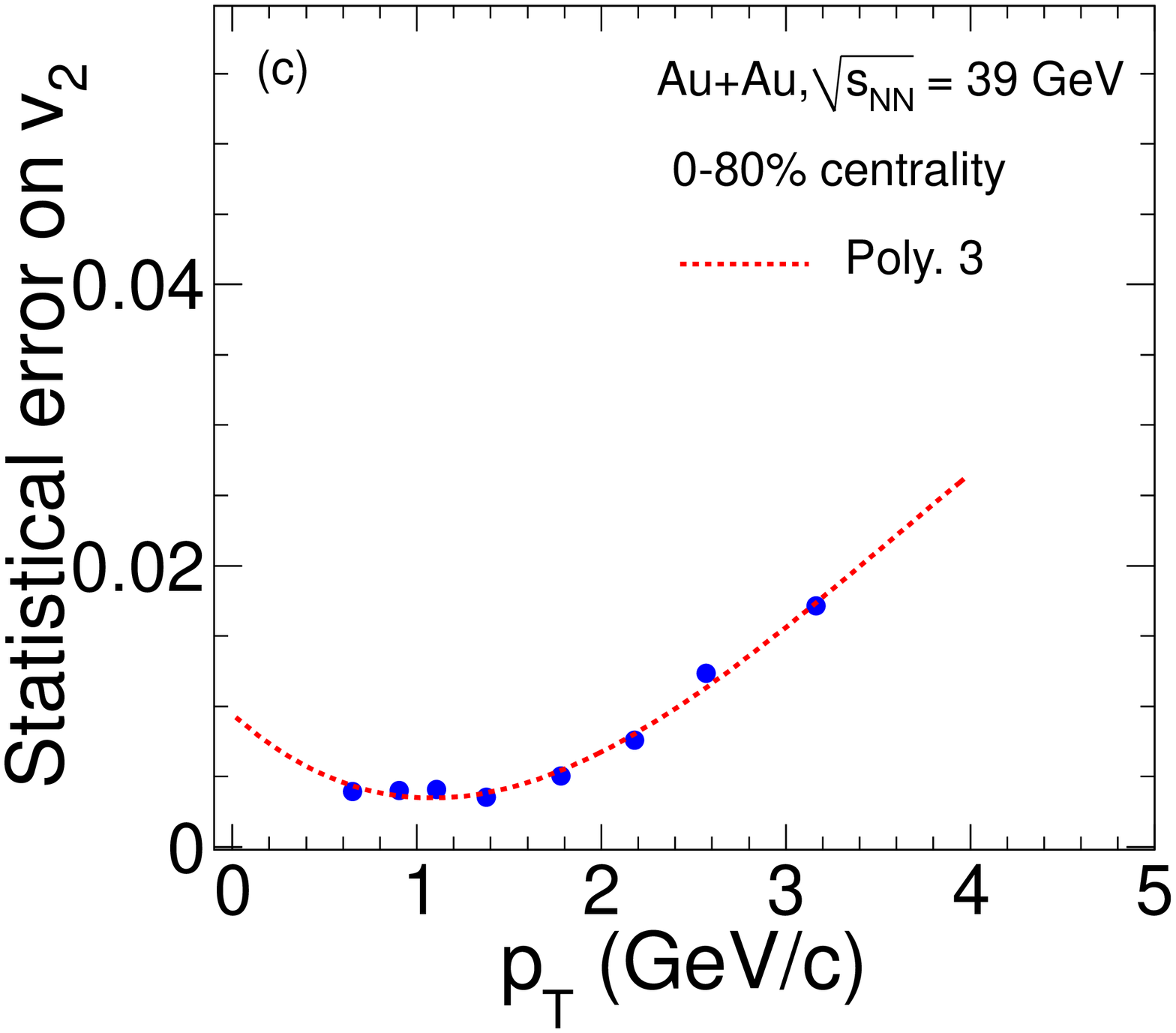}
\caption{(Color online) Panel (a): The $\phi$-meson $v_{2}(p_{T})$ at
  $\sqrt{s_{NN}}$ =39 GeV for 0-80$\%$ centrality bin is fitted with
  $3^{rd}$ order polynomial (Poly. 3) and with function described in
  Eq.~\ref{ncq_fun}.
Panel (b): The preliminary $\phi$-meson $dN/dp_{T}$ vs $p_{T}$ at
$\sqrt{s_{NN}}$ =39 GeV for 0-80$\%$ centrality bin is fitted with
levy function.
Panel (c): Statistical error on $v_{2}(p_{T})$ at
  $\sqrt{s_{NN}}$ =39 GeV for 0-80$\%$ centrality bin is fitted with
  $3^{rd}$ order
  polynomial. 
}
\label{avg_v2_method}
\end{center}
\end{figure}
 The $p_{T}$-averaged $v_{2}$ may have a better statistical precision than the $p_{T}$ differential measurements. To calculate
the $\langle v_{2}\rangle$ of  $\phi$ mesons, each $v_{2}(p_{T})$
distribution was fitted with function (shown in  Fig.~\ref{avg_v2_method}): a $3^{rd}$ order polynomial
function and a function of the form
\begin{equation}
  f_{v_{2}}(n)  = \frac{an}{1+exp[-(p_{T}/n -b)/c]} -dn,
\label{ncq_fun}
\end{equation}
where $a$, $b$, $c$ and $d$ are free parameters and $n$ is the number of
constituent quarks.  This function was inspired by parameterizations of quark
number scaling~\cite{ncq_fit_xin}. The $p_{T}$ distribution of
 $\phi$ mesons has been fitted with Levy function as shown in 
 panel (b) of Fig.~\ref{avg_v2_method}. The functional form of Levy function is
given by
\begin{equation}
\begin{array}{ll}
 f_{Levy}(p_{T})=  & \\\frac{dN}{dy}\times\frac{(n-1)(n-2)}{2\pi n
   T(nT+m_{0}(n-2))}\times(1+\frac{\sqrt{p_{T}^{2}+m_{0}^{2}}-m_{0}}{nT})^{-n},
\end{array}
\end{equation}
where $T$ is known as the inverse slope parameter, $dN/dy$ is the $\phi$-meson yield
per unit rapidity, $m_{0}$ is the rest mass of $\phi$ meson and $n$ is the Levy function
parameter.
The $\langle
v_{2}\rangle$  for each choice of $v_{2}(p_{T})$ parameterization is given by the
integral of the corresponding distributions normalized by integral of
the $p_{T}$ distribution. In addition, the $\langle v_{2}\rangle$ has
been calculated directly from measured data points of $v_{2}(p_{T})$
with corresponding yield obtained from the fit function to the $p_{T}$ distribution. The final $\langle v_{2}\rangle$ was
obtained by calculating the mean of the three $\langle v_{2}\rangle$
results and the systematic error was estimated from maximum deviation
from the mean value. There are two sources for the statistical error, one is error on  $p_{T}$
distribution and other is error on $v_{2}(p_{T})$. Since the error on
$dN/dp_{T}$ is very small compared to that on $v_{2}(p_{T})$, one can
simply neglect the error of $dN/dp_{T}$. Hence, only  errors on $v_{2}(p_{T})$
are taken care for calculation of  final statistical error on $\langle
v_{2}\rangle$. The errors on $v_{2}$ are parameterized as a function of $p_{T}$ and extrapolated
to low and high $p_{T}$ as shown in panel (c) of
Fig.~\ref{avg_v2_method}. Figure~\ref{avg_v2_method} is repeated for all the energies studied.
 For $\langle v_{2}\rangle$ calculation in data, the final
$\phi$ mesons spectra  and $v_{2}(p_{T})$  at 62.4 and 200 GeV published by STAR has been
used~\cite{phi_200_prc}. For the other energies, the STAR preliminary spectra~\cite{xp_QM} and final
$v_{2}(p_{T})$~\cite{v2_BES_prc} has been used. \\ 
The $p_T$ integrated $\phi$-meson $v_{2}$ for Au+Au
 minimum-bias collisions at mid-rapidity($|\it y|$ $<$ 1.0 ) are  compared to the corresponding
AMPT model calculation at various beam energies in
Fig.~\ref{avg_v2}. In contrast to observations from the data, the
$\langle v_{2}\rangle$ values from model remain constant for all the
energies for a given parton-parton interaction cross-section. 
The $\langle v_{2}\rangle$ of $\phi$ mesons for  $\sqrt{s_{NN}}$
$\geq$ 19.6 can be explained by the AMPT with string melting depending on parton-parton cross-section.
The AMPT-SM model with 10mb parton-parton cross-section explain the
data very well at $\sqrt{s_{NN}}$ =62.4 and 200 GeV, where as 3mb parton-parton cross-section is sufficient
to describe the data at  $\sqrt{s_{NN}}$ = 19.6, 27 and 39 GeV. On the
other hand, both the AMPT-SM and AMPT Default model over predict  data
at $\sqrt{s_{NN}}$ = 11.5 GeV, indicating negligible contribution of
the partonic collectivity to the final collectivity. Due to very small
statistics at $\sqrt{s_{NN}}$ = 7.7 GeV, $\langle v_{2}\rangle$ are
not shown here.
The observation that different parton-parton cross sections are needed to
 explain the data within the transport model framework indicates that 
 the $\eta$/$s$ changes with beam energy. Higher the cross section, smaller
 is the $\eta$/$s$ expected for the system. This qualitative observation of variation 
in the value of $\eta$/$s$ with beam energy is consistent with the expectations from various 
calculations as reported in~\cite{eta_by_s}.

From the Fig.~\ref{avg_v2} , one can conclude that as the energy decrease
contribution to the collectivity from the partonic phase also decreases and for
 $\sqrt{s_{NN}}$ $\leq$ 11.5 GeV, the hadronic interaction plays a
 dominant role in experimentally observed data.
\\
\section{Summary and Conclusion}
In summary, a compilation of the available data for elliptic flow of
$\phi$ mesons has been presented.  The implications of these results on the quark-hadron phase
transition has been discussed by comparing experimental data with
AMPT model. The AMPT model with string melting scenario
quantitively explain the data at  $\sqrt{s_{NN}}$ $\geq$ 19.6 GeV  by
varying parton-parton interaction cross-section from 3mb to 10mb.
 The
AMPT Default model under predict that experimental data for
$\sqrt{s_{NN}}$ $\geq$ 19.6 GeV. This tells that there is a substantial
contribution of partonic collectivity to the final collectivity for $\sqrt{s_{NN}}$ $\geq$ 19.6 GeV.
However, both the AMPT-SM and AMPT Default can not explain the trend of
$\phi$-meson $v_{2}$ as function of $p_{T}$ at  $\sqrt{s_{NN}}$ = 7.7
and 11.5 GeV. Also the $\langle v_{2}\rangle$ from AMPT default
over-predicts the data at  $\sqrt{s_{NN}}$ = 11.5 GeV. This indicates
that possible turn off of partonic interaction starts at $\sqrt{s_{NN}}$ $\leq$
11.5 GeV. Due to large statistical error on $\phi$ $v_2$ at
$\sqrt{s_{NN}}$ = 7.7 GeV, it is not possible to make any conclusions.
The comparison of the experimental data on the beam energy dependence
of the average elliptic flow of $\phi$ meson with the corresponding results from a transport model calculation with varying parton-parton cross section suggests that the partonic contribution to the collectivity decreases and possibly the value of the $\eta$/$s$ of the system increases as the beam energy decreases.
The $\phi$-meson $v_{2}$ measurement should be one of the main focuses
in the proposed BES phase II program and also in FAIR experiment at
GSI to explore the phase diagram further.\\
\noindent{\bf Acknowledgments}\\
I thank Dr. Bedangadas Mohanty for useful discussions and help in the preparation of the manuscript. 
Financial support from DST SwarnaJanti project, Government of India is gratefully acknowledged.


\normalsize
\begin{thebibliography}{99}

\bibitem{whitepapers} I. Arsene  {\it et al.}  (BRAHMS Collaboration), Nucl. Phys. \textbf{A 757}, 1 (2005);
                      B. B. Back  {\it et al.} (PHOBOS Collaboration), Nucl. Phys. \textbf{A 757}, 28 (2005);
                      J. Adams  {\it et al.}   (STAR Collaboration),   Nucl. Phys. \textbf{A 757}, 102 (2005);
                      K. Adcox  {\it et al.}   (PHENIX Collaboration), Nucl. Phys. \textbf{A 757}, 184 (2005).


\bibitem{bes_res1} L. Adamczyk  {\it et al.} (STAR Collaboration),
  Phys. Rev. \textbf{C 86}, 054908 (2012).
\bibitem{v2_BES_prc} L. Adamczyk {\it et al.} (STAR Collaboration),
  Phys. Rev. \textbf{C 88}, 014902 (2013).
\bibitem{v2_BES_prl} L. Adamczyk {\it et al.} (STAR Collaboration),
  Phys. Rev. Lett. \textbf{110}, 142301 (2013).


\bibitem{bes_moti} B. I. Abelev  {\it et al.} (STAR Collaboration), Phys. Rev. \textbf{C 81},
024911 (2010).


\bibitem{hydro} P.F. Kolb {\it et al.} Nucl. Phys. \textbf{A 715},
  653c (2003).
\bibitem{hydro1} D. Teaney {\it et al.} Phys. Rev. Lett. \textbf{86}, 4783 (2001). 
\bibitem{hydro2} P. F. Kolb and U. Heinz, arXiv:[nucl-th/0305084]. 
\bibitem{hydro3}  P. F. Kolb {\it et al.} Phys. Lett. \textbf{B 500}, 232 (2001). 
\bibitem{hydro4} H. Sorge, Phys. Rev. Lett. \textbf{78}, 2309 (1997). 


\bibitem{method} A. M. Poskanzer and S. A. Voloshin,  Phys. Rev. \textbf{C 58},
  1671  (1998).

\bibitem{early_v2} B. Zhang  {\it et al.} Phys. Lett. \textbf{B 455}, 45 (1999).

\bibitem{ncq1} D. Molnar and S. A. Voloshin, Phys. Rev. Lett. \textbf{91},
  092301  (2003). 
\bibitem{ncq1a} J. Adams {\it et al.} ( STAR Collaboration)
  Phys. Rev. Lett. \textbf{92}, 052302 (2004).


\bibitem{ncq_phi} C. Nonaka {\it et al.} Phys. Lett.  \textbf{ B 583},
  73 (2004); V. Greco {\it et al.} Phys. Rev. \textbf{ C 68}, 034904
  (2003);  R. J. Fries {\it et al.}
  Ann. Rev. Nucl. Part. Sci. \textbf{ 58}, 177 (2008).


\bibitem{ncq2} B. I. Abelev {\it et al.} (STAR Collaboration)
  Phys. Rev. Lett. \textbf{99}, 112301 (2007). 


\bibitem{ncq_urqmd1} P. Bhaduri and S. Chattopadhyay Phys. Rev. \textbf{C 81}, 034906 (2010).
\bibitem{ncq_urqmd2} K. J. Wu {\it et al.}  Nuclear Physics \textbf{A
    834}, 303c (2010).

\bibitem{ncq_urqmd3} M. Bleicher and X. Zhu, Eur. Phys. J. \textbf {C
    49}, 303 (2007).

\bibitem{smallx} A. Shor, Phys. Rev. Lett. \textbf{54}, 1122 (1985) and  J. Rafelski and B. Muller, Phys. Rev. Lett. \textbf{48}, 
  1066 (1982).
\bibitem{NBN}  Md. Nasim, B. Mohanty and N. Xu , Phys. Rev. \textbf{C 87}, 014903 (2013).
\bibitem{BN}  B. Mohanty and N. Xu , J. Phys. \textbf{G 36}, 064022  (2009).
\bibitem{ampt} Zi-Wei Lin and C. M. Ko, 
               Phys. Rev. \textbf{C 65}, 034904 (2002);\\
               Zi-Wei Lin {\it et al.},
               Phys. Rev. \textbf{C 72}, 064901 (2005);\\
               Lie-Wen Chen {\it et al.}, 
               Phys. Lett. \textbf{B 605} 95 (2005).
\bibitem{hijing}X. N. Wang and M. Gyulassy, Phys. Rev. \textbf{D 44},
  3501 (1991).
\bibitem{ZPC} B. Zhang, Comput. Phys. Commun. \textbf{109}, 193 (1998).
\bibitem{lund} B. Andersson {\it et al.}  Phys. Rep. \textbf{97},31 (1983).


\bibitem{ncq_fit_xin} X. Dong {\it et al.} Phys. Lett. \textbf{B 597},
  328–332 (2004).
\bibitem{phi_200_prc} B.I. Abelev  {\it et al.} (STAR Collaboration), Phys. Rev. \textbf{C 79}, 064903 (2009).
\bibitem{xp_QM} Talk of Xiaoping Zhang (for the STAR collaboration) at Quark Matter 2012.\\
                           Talk of Md. Nasim (for the STAR collaboration) at CPOD 2103.\\
                           Talk of Md. Nasim (for the STAR
                           collaboration) at  SQM 2103.


\bibitem{eta_by_s} S. Plumari {\it et al.} arXiv:1304.6566 [nucl-th];
  R. A. Lacey   {\it et al.} arXiv:1305.3341 [nucl-ex];  L. P. Csernai
  {\it et al.} Phys. Rev. Lett. {\bf 97}, 152303 (2006); R. A. Lacey
  {\it et al.} Phys. Rev. Lett. {\bf 98}, 092301 (2007).



\end{thebibliography}
\end{document}